# Soliton Microcombs in Integrated Chalcogenide Microresonators

Di Xia[1,5], Zelin Yang[1,5], Pingyang Zeng[1,5], Bin Zhang[1,2]*, Jiayue Wu[1], Zifu Wang[1], Jiaxin Zhao[1], Mingqi Gao[1], Yufei Huang[1], Jianteng Huang[1], Liyang Luo[1], Dong Liu[1], Shuixian Yang[1], Hairun Guo[3], and Zhaohui Li[1,2,4,]*

[1] Guangdong Provincial Key Laboratory of Optoelectronic Information Processing Chips and Systems, School of Electrical and Information Technology, Sun Yat-sen University, Guangzhou, 510275, China.
[2] Key Laboratory of Optoelectronic Materials and Technologies, Sun Yat-sen University, Guangzhou, 510275, China.
[3] Key Laboratory of Specialty Fiber Optics and Optical Access Networks, Shanghai University, Shanghai, 200444, China.
[4] Southern Marine Science and Engineering Guangdong Laboratory (Zhuhai), Zhuhai, 519000, China.
[5] These authors contributed equally: Di Xia, Zelin Yang, Pingyang Zeng

*e-mail: zhangbin5@mail.sysu.edu.cn, lzhh88@mail.sysu.edu.cn

## Abstract：

Photonic integrated microcombs have enabled advanced applications in optical communication, microwave synthesis, and optical metrology, which in nature unveil an optical dissipative soliton pattern under cavity-enhanced nonlinear processes. The most decisive factor of microcombs lies in the photonic material platforms, where materials with high nonlinearity and in capacity of high-quality chip integration are highly demanded. In this work, we present a home-developed chalcogenide glasses-$Ge_{25}Sb_{10}S_{65}$ (GeSbS) for the nonlinear photonic integration and for the dissipative soliton microcomb generation. Compared with the current integrated nonlinear platforms, the GeSbS features wider transparency from the visible to 11-μm region, stronger nonlinearity, and lower thermo-refractive coefficient, and is CMOS compatible in fabrication. In this platform, we achieve chip-integrated optical microresonators with a quality (Q) factor above $2\times10^6$, and carry out lithographically controlled dispersion engineering. In particular, we demonstrate that both a bright soliton-based microcomb and a dark-pulsed comb are generated in a single microresonator, in its separated fundamental polarized mode families under different dispersion regimes. The overall pumping power is on the ten-milliwatt level, determined by both the high Q-factor and the high material nonlinearity of the microresonator. Our results may contribute to the field of nonlinear photonics with an alternative material platform for highly compact and high-intensity nonlinear interactions, while on the application aspect, contribute to the development of soliton microcombs at low operation power, which is potentially required for monolithically integrated optical frequency combs.

# Introduction

Integrated nonlinear photonics have combined nonlinear optics with state-of-the-art photonic integration and have unprecedentedly promoted light-matter interactions in terms of efficiency, bandwidth, and coherence[1-6]. They have enabled revolutionary techniques including chip-integrated optical frequency combs (OFCs)[5, 7], ultra-high bandwidth electro-optical modulation[2], and chip-scale quantum optics[8]. Of particular interest is the dissipative soliton-based OFCs in photonic integrated microresonators, which can be generated at a low pump power[9] and have a broad bandwidth with fully coherent laser lines, benefiting from cavity-enhanced nonlinear efficiency and the lithographically controlled dispersion engingeering[7]. To date, they have enabled various advanced applications, including massive parallel optical telecommunication[10-12], low-noise microwave synthesis[13], parallel LiDAR[14, 15], photonic neuromorphic computing[16], and other photonic functionalities[3, 17, 18] for chip-scale time and frequency measurements.

Tremendous nonlinear material platforms have been developed to realize high-performance soliton microcombs[7]. Recent advances have witnessed several material platforms that could successfully support integrated soliton microcombs, such as silicon nitride ($Si_3N_4$)[6, 19, 20], high-index doped silica (Hydex)[11, 21], aluminum nitride (AlN)[22, 23], gallium nitride (GaN)[24], lithium niobate ($LiNbO_3$)[25-27], silicon carbide (SiC)[8, 28] as well as AlGaAs[9, 29] on insulator. Amorphous $Si_3N_4$ seems to be particularly fruitful in soliton combs, which has wide transparency, free from two-photon absorption (TPA), higher nonlinearity ($2.4 \times 10^{-19}$ $m^2/W$) than Hydex, and has supported record-high quality (Q) factors in integrated microresonators[6]. On another aspect, semiconductor platforms such as AlGaAs on insulator could introduce even higher nonlinearity to reduce the pump power of soliton combs to sub-milliwatt (mW) level[9]. Crystalline-based platforms including AlN and LiNbO$_3$ introduce extra quadratic nonlinearities to generate the ultra-broadband soliton combs, which are potentially required for the self-referencing of microcombs[22, 30]. Particularly, most recently, monolithic microcomb chips have been developed based on the hybrid integration of such high-Q microresonators with III-V lasing chips, which are prone to be fully integrated in realistic systems[31, 32].

Yet, limitations are also realized in operating such platforms for soliton combs with low pump power, compact and flexibility in high volume microcomb-chip fabrication. For instance, additional processes in the fabrication of Si$_3$N$_4$ thick films for bright soliton microcombs are required, including crack-mitigation, chemical-mechanical polishing, and high temperature anneals (exceeding 1100 ºC) strategies[33, 34]. Crystalline platforms require the wafer-bonding process to be integrated on insulator substrates. In this way, the flexibility of performing dispersion engineering is reduced, and the yield of devices may be degraded as well[6, 35]. Additionally, semiconductor resonators usually feature a high thermo-refractive coefficient (TOC), which may hinder the existence of soliton combs[24, 29]. As such, there exists a continuous motivation of seeking advanced material platforms, which could patch up the above issues and can be alternative to present platforms. In the meantime, while the current soliton microcombs are mostly designed and operated in the telecom band, it is equally essential for the material to open access in new wavelength regions for a broader range of microcomb applications, such as precise spectroscopy in the mid-infrared (MIR)[36].

Here, we demonstrate a modified chalcogenide glasses (ChGs)[4, 37, 38] based photonic platform for soliton microcomb generation in high-quality and chip-scale microresonators. The material is $Ge_{25}Sb_{10}S_{65}$ (GeSbS) that inherits the ultrabroad transmission window with absence of TPA, large refractive index and Kerr nonlinearity[37-42] of ChGs, and shows flexibility in photonic integration on silicon-based chips[43-45]. In the meantime, with modified compounds, GeSbS could overcome existing problems of As-based ChGs, and features an improved laser damage threshold (LDT) with a reduced TOC. We fabricate chip-integrated GeSbS microresonators with an intrinsic Q-factor above $2\times10^6$, and demonstrate the generation of both the bright dissipative soliton microcombs and the dark pulse microcombs. In particular, the two microcomb regimes can be implemented in a single microresonator in the two fundamental polarized mode families that are lithographically engineered to induce normal and anomalous dispersive effects. Given a high Q-factor and high nonlinearity of the microresonator, the soliton comb is supported with a low pump power on the 10-mW level, which is prone to implementing monolithic frequency comb chips for integrated metrology applications, such as compact dual-comb spectrometers and MIR spectroscopy.

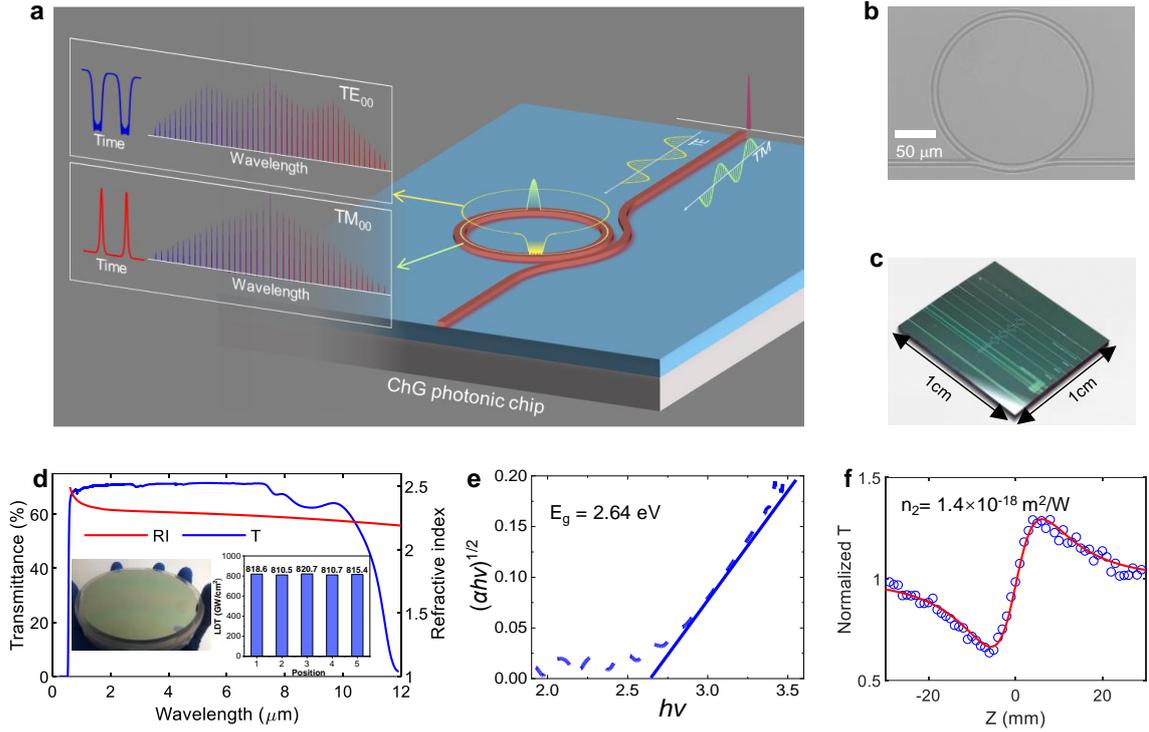

**Fig.1 Nonlinear Chalcogenide photonic chip for soliton microcomb.** (a) Schematic illustration of soliton microcomb generations in an integrated ChG microresonator, including dark and bright soliton combs in the $TE_{00}$ and $TM_{00}$ modes. (b) Scanning electron microscopy (SEM) image of a GeSbS microresonator with a radius of 100 μm with an integrated pulley bus waveguide. (c) Photograph of a fabricated 1 cm×1 cm GeSbS photonic chip comprises several microresonators with different structural parameters. (d) Measured transparency window and refractive index of the GeSbS bulk material. Inset: photograph of the GeSbS film (left) and the measured LDT at 5 different positions of the film using the femtosecond pulse laser(see materials and methods part)r, respectively (right). (e) Measured Tauc's plot for bandgap determination of GeSbS film. (f) Z-scan (close aperture) trace for nonlinear refractive index determination of GeSbS film.

## Results

**Nonlinear GeSbS photonics for soliton microcomb generation.**

  **Figure 1a** illustrates the schematic of ChG photonic chip-based soliton microcomb generation. With precise dispersion engineering on a high-Q GeSbS microresonator, both a bright soliton-based microcomb and a dark pulse-based comb can be generated in the same microresonator in different fundamental polarized mode families. In detail, the bright soliton comb is in the $TM_{00}$ mode family that shows anomalous Group velocity dispersion (GVD), and the dark pulse comb is in the TE$_{00}$ mode family with normal GVD. The material platform is amorphous GeSbS (Ge$_{25}$Sb$_{10}$S$_{65}$), which can directly adhere to the crystal substrates by thermal evaporation at a low temperature (less than 350 °C)[24]. The material also features a high laser damage threshold (ca. 820.7 GW/cm$^2$), strong Kerr nonlinearity (1.4×10$^{-18}$ m$^2$/W at 1550 nm)[46], large linear index (n$_0$~2.2), large bandgap (2.64 eV), ultrabroad transparency (from 0.5 μm to up to 11 μm, no TPA), and a relatively low thermo-optic coefficient (ca. 3.1 × 10$^{-5}$/K, see **Fig. 4a**).

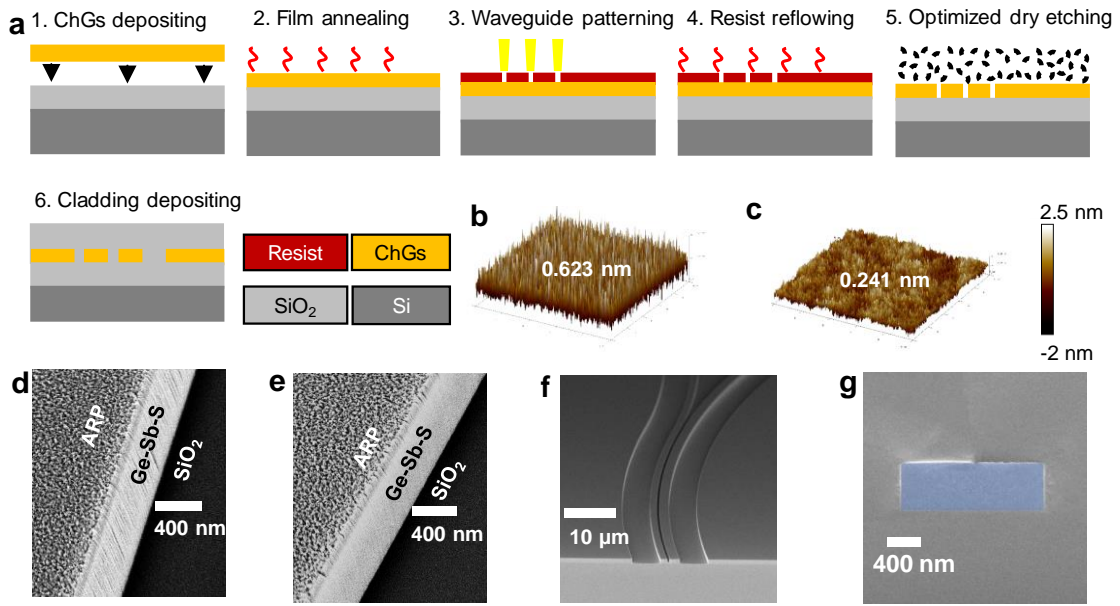

**Fig.2 Fabrication of high Q-factor GeSbS microresonators. (a)** Fabrication processes for achieving high Q integrated microresonators. **(b), (c)** 3D Atomic-force microscopy (AFM) images of GeSbS film in 5 × 5 μm$^2$ area on-chip before and after the annealing process at 350 °C. **(d-g)** SEM images of the sidewall of the waveguides before and after the improved dry etching process, the pulley coupling structure, and the cross-section with a silica cladding waveguide, respectively. In (f), a compact photonic chip was demonstrated with the optimized coupling efficiency of -3.5 dB (45 %) per facet at 1550 nm using inverse taper couplers at the chip edge.

**Fabrication of high *Q*-factor GeSbS microresonators.**

The lack of sufficiently high Q in ChG platforms has limited the ability to harness their dramatic material properties for nonlinear optics applications[9]. The surface roughness is critical for achieving high $Q$-factor photonic integrated devices[34]. An improved waveguide fabrication process was provided to minimize the surface roughness of the top and sidewall of the microresonators, including optimized thermal-annealing of ChG film, thermal-refluxing the patterned mask-layer, transferring the pattern by an optimized plasma etching process, see **Fig. 2a**. The thermal annealing process is an effective and straightforward method to remove the surface roughness of amorphous GeSbS film. The root-mean-squared (RMS) roughness is decreased from 0.62 nm of the as-deposited film to 0.24 nm of the film annealed at 350 °C correspondingly. Moreover, we focused on reducing scattering loss from *in-situ* polymer formation on the sidewall of waveguides by optimizing the dry etching process[16]. An optimized $CF_4/CHF_3/Ar$ gas-based reactive ion etching and inductively coupled plasma (RIE-ICP) etching recipe was successfully achieved for smoother sidewalls of the microresonator by adjusting the flow of $O_2$ and $CF_4$ gases[34, 46], see **Fig. 2d and 2e**. Based on the abovementioned processes, the top and sidewall surfaces of the microresonator could still be smooth after the resist removing, see **Fig. 2f**. As a result, a $Q$-factor of more than $10^6$ was achieved, in contrast to the $Q$-factor of GeSbS microresonator with the same geometric dimensioning using the previously reported RIE-ICP etching recipe was only $10^5$ level[47]. Moreover, the cross-section of the waveguide presents an almost vertical sidewall (see **Fig. 2g**), which is beneficial for accurate geometry dispersion control.

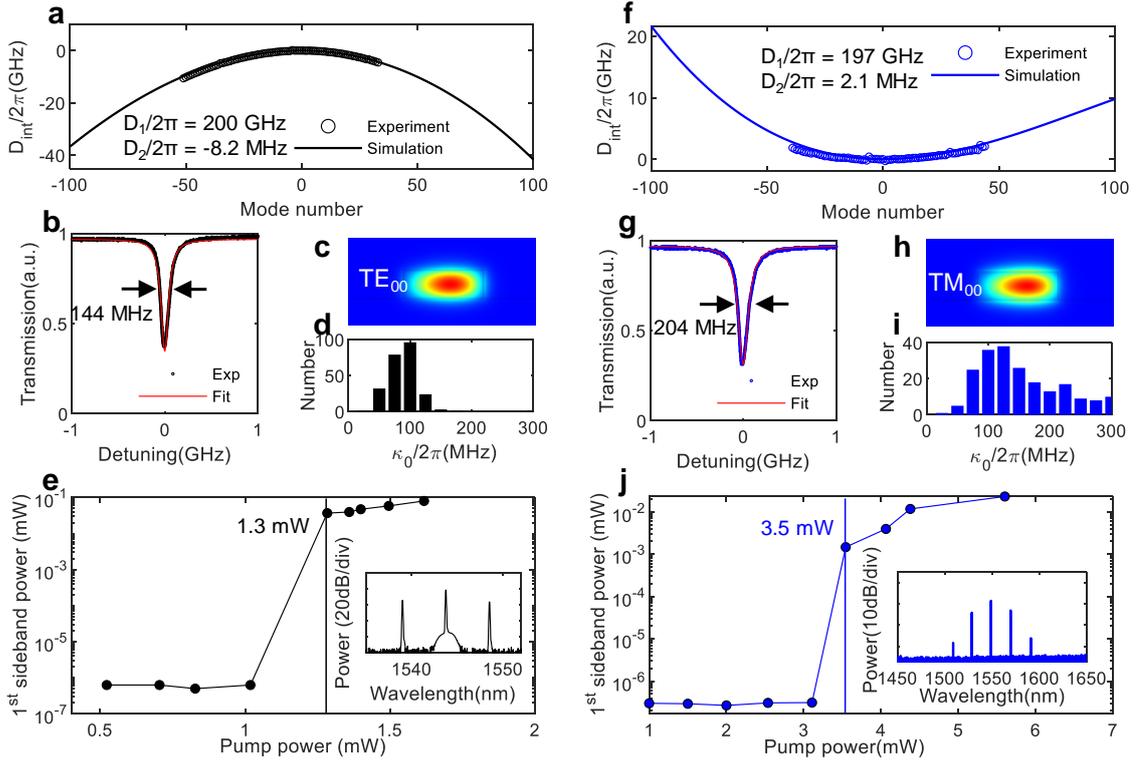

**Fig.3 Characterization of high *Q*-factor GeSbS microresonators with TE$_{00}$ resonances (a-e) and TM$_{00}$ resonances (f - j), respectively. (a), (f)** The calculated and measured resonator dispersion for TE$_{00}$ and TM$_{00}$ modes with a radius of 100 μm, and the cross-section is 2.4 μm × 0.8 μm (width × height), respectively. The measured FSR ($D_1/2\pi$) is 200 GHz, 197 GHz, fitted GVD ($D_2/2\pi$) is -8.2 MHz, 2.1 MHz for TE$_{00}$, TM$_{00}$ mode, respectively. **(b), (g)** The measured transmission spectra at 1556 nm and 1555 nm, revealing full-width-at-half-maximum (FWHM) to be 144 MHz and 204 MHz from the Lorentz fitting curves. **(c), (h)** the corresponding mode profiles. **(d), (i)** Histogram of intrinsic linewidths. **(e), (j)** The output power of the first generated FWM sideband as a function of input power. Insets: the measured output optical spectra, showing 3-FSR (FSR = 200 GHz) and 16-FSR (FSR = 197 GHz) spaced 1$^{st}$ sidebands for or TE$_{00}$ and TM$_{00}$ modes, respectively.

**Characterization of high *Q*-factor GeSbS microresonators.**

Next, we systematically studied the dispersion, Q-factors, and OPO threshold powers based on GeSbS microresonators[48]. To further ensure accurate dispersion control of the waveguide, refractive index variation during the annealing process was also considered in device design, as shown in SI **Fig. S1**. Normal material GVD was exhibited in the telecom band (ca. -400 ps/nm/km at 1550 nm) by measuring the linear indices of GeSbS film, which had to tailor the waveguide dimensions to achieve strongly anomalous geometric GVD. In the context of microresonators, the dispersion can lead to a deviation of resonances from equidistant mode spacing. The integrated dispersion $D_{int}$ can express[49]:

$$D_{int} = \omega_\mu - \omega_0 - D_1\mu = \frac{D_2\mu^2}{2!} + \frac{D_3\mu^3}{3!} + \sum_{m>3}\frac{D_m\mu^m}{m!}$$

where μ, $\omega_\mu$, and m are the relative mode numbers, angular frequencies of the resonances, and a positive integer, respectively. We experimentally measured the frequency difference using a fiber ring resonator and a fiber loop[21]. B*y* analyzing the resonance variation from equidistant mode spacing, the $D_2/2\pi$ of $TE_{00}$ and $TM_{00}$ mode are -8.2 MHz and 2.1 MHz at the center frequency of around 193.4 THz, see **Fig. 3a and 3f.** Considering the material dispersion of GeSbS and $SiO_2$, the microresonator dispersion as a function of relative mode numbers was also calculated by means of the finite element method (see Materials and methods). The simulated integrated dispersion profiles for both the $TE_{00}$ and the $TM_{00}$ mode families are shown in **Fig. 3a and 3f**, which have a good agreement with the experimental measurements. The fluctuations of the thickness (ca. ±6.5 nm) and refractive index (ca. ±0.0075) of the 4-inch ChG film are also considered in precise GVD control (see **SI Fig. S1**).

The intrinsic *Q*-factor of our microresonators is typically above $10^6$ for both mode families, see **Fig. 3b and 3g**. In a number of microresonator samples on the same fabrication batch, all the resonances within the measurement range (1510-1630 nm) are characterized, which show a mean value of *Q*-factor of ca. $1.97\times10^6$ for the $TE_{00}$ mode family and ca. $1.38\times10^6$ for $TM_{00}$ mode family, see **Fig. 3d and 3i.** The pump threshold for OPO in such high-Q microresonators was also characterized by measuring the output powers of the first generated FWM sidebands with different input power, which for a pair of selected $TE_{00}$ and TM$_{00}$ modes are 1.3 mW and 3.5 mW, respectively. Moreover, in a 20 μm-radius microresonator, the measured threshold power can be as low as 0.78 ± 0.1 mW (see SI **Fig. S3** ), as it is scaled by the free spectral range (FSR) of the resonator:

$$P_{th} = \frac{\pi}{8} \frac{n}{n_2} \frac{\nu_0}{\nu_{FSR}} \frac{A_{eff}}{Q_i^2} \times \frac{(1+\kappa)^3}{\kappa} \qquad (2)$$

where n is the linear, n$_2$ is the nonlinear refractive index, $\nu_0$ is the pump frequency $\nu_{FSR}$ is FSR of the resonator, A$_{eff}$ is the effective mode area of the resonator, the coupling factor $\kappa = \kappa_{ex}/\kappa_i = Q_i/Q_c$, $\kappa_{ex}$ is the coupling rate, and $\kappa_i$ is the intrinsic rate of the resonator, respectively. Given that n$^2$ has been experimentally measured to be $1.3 \times 10^{-18}$ m$^2$/W (which is almost five times higher than that of Si$_3$N$_4$), we noticed that the calculated threshold power is in good agreement with the measured value, namely the estimated power is 0.72 mW (TE$_{00}$, 20 μm-radius), 1.4 mW (TE$_{00}$, 100 μm-radius) and 3.4 mW (TM$_{00}$, 100 μm-radius), respectively.

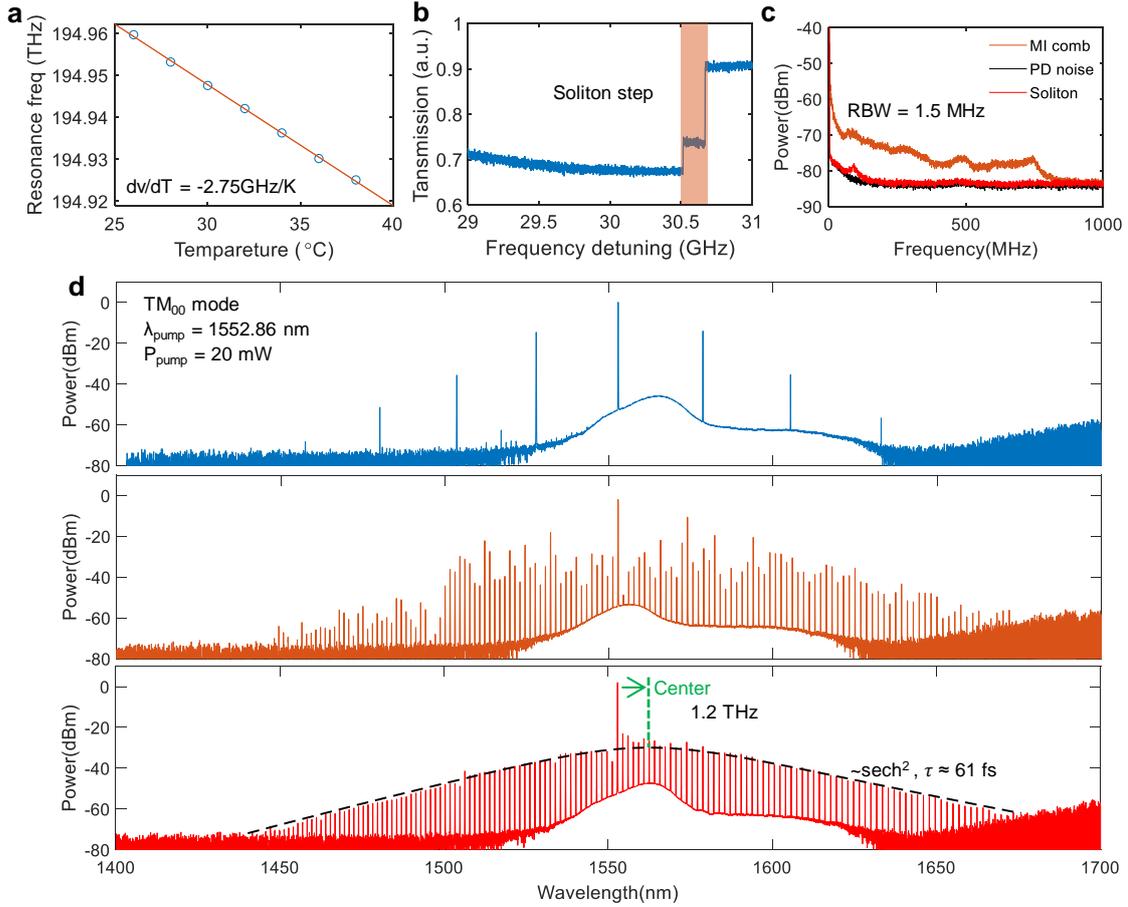

**Fig.4 Bright soliton comb generation in a GeSbS microresonator. (a)** Measured resonator frequency shift versus temperature for determining TOC of GeSbS resonators. **(b)** Transmission spectrum of the resonance when a laser is swept. The "soliton step" can be observed, indicating a transition to soliton state. **(c)** Intensity noise spectrum of MI comb state and soliton states displayed in an electrical spectrum analyzer (ESA). **(d)** Mode-locked bright soliton comb generation as the pump laser is red-detuned into the cavity resonance (top to bottom), the pump power is fixed at ca. 20 mW. The single soliton was fitted with a sech$^2$ envelope (dashed dark line), indicating an estimated duration of 61 fs. The green arrow shows Raman-induced redshift of the soliton spectrum concerning the pump line.

**Generation of soliton microcombs in GeSbS microresonators**

We measured the mode-locked soliton comb operations to demonstrate the advance of our GeSbS platform for microcomb formations. Because high thermo-optical instability is challenging for accessing soliton in resonators[35, 50], a lower TOC (dn/dT, where T is temperature) is beneficial. We measured the thermal-optic shift of the resonance frequency when heating the entire chip, see **Fig. 4a**. From the fitting of temperature-frequency data, we determined TOC utilizing the equation:

$$\frac{dn}{dT} \approx -\frac{n}{\upsilon}\frac{d\upsilon}{dT} \approx 3.1\times10^{-5} \text{ K}^{-1} \quad (3)$$

Such a TOC value is comparable to that of $Si_3N_4$ ($2.4\times10^{-5}$ K$^{-1}$) and is around one order of magnitude lower than that of AlGaAs ($3.6\times10^{-4}$ K$^{-1}$) and GaN (ca. $10^{-4}$ K$^{-1}$). We applied the laser tuning method to form the soliton microcomb. The experimental setup for soliton generation and characterization is shown in SI **Fig. S4**. During the laser scan process, primary comb separated by 16-FSR (FSR = 197 GHz) and modulation instability (MI) combs were first observed when the pump light was located at the blue-detuned side of the resonance (see **Fig. 4d**, top and middle). A "soliton step" was observed when the laser was tuned to the red-detuned side, indicating the formation of the dissipative solitons in the cavity[51], see **Fig. 4b**. By stopping the laser frequency on the step and slightly adjusting the detuning, a single-soliton-state microcomb was observed with a wavelength span from 1440 nm to 1680 nm (see **Fig. 4d**, bottom). Simultaneously, the drastic reduction of intensity noise of output light shows the transition to a soliton regime with low noise (see **Fig. 4c**). The soliton duration can be estimated from its 3-dB bandwidth as 61 fs by fitting the final spectrum of a single soliton state with a sech$^2$ envelope. We also observed the Raman effect-induced soliton red spectral shift (ca. 1.2 THz in the present case) concerning the pump line (the green arrow in **Fig. 4d**)[52]. As a result, the lower TOC coefficient and low comb operation power persuade the stable soliton microcombs generations in our GeSbS microresonator without complex laser tuning schemes or auxiliary lasers[24].

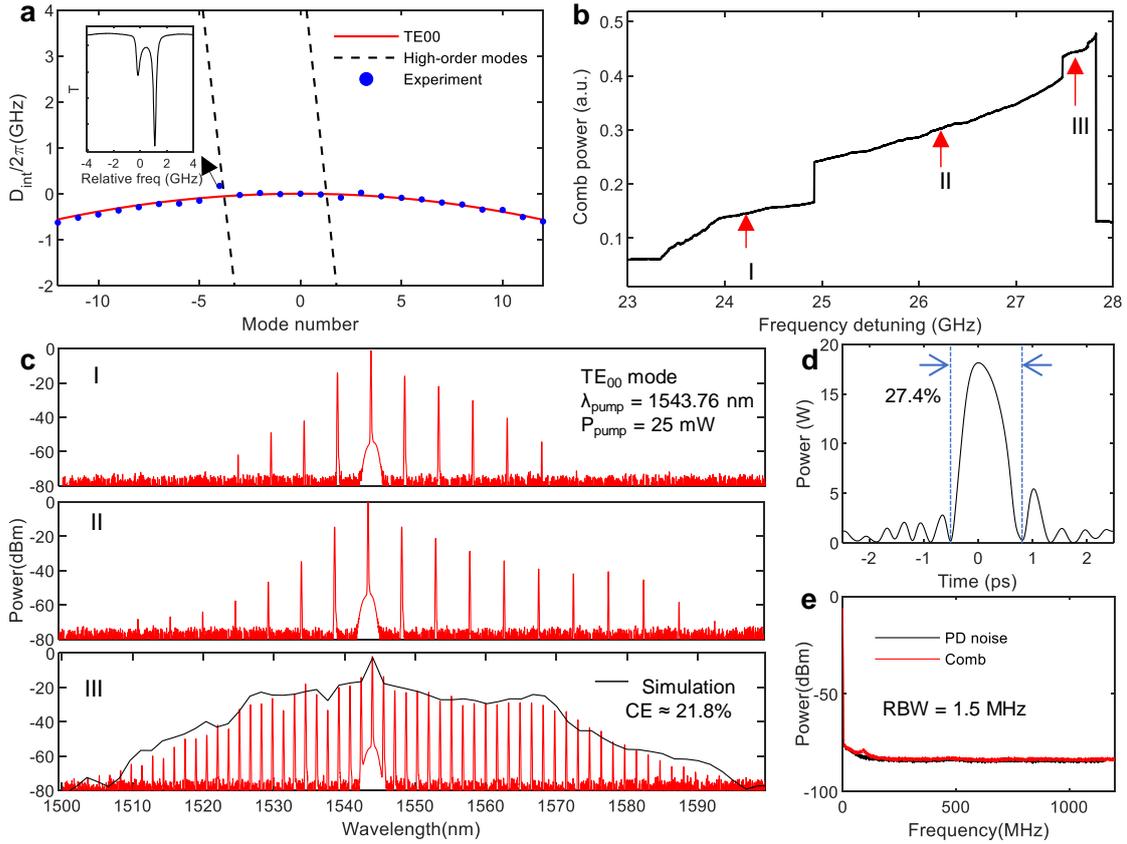

**Fig.5 Dark-pulse comb generation in a GeSbS microresonator.** (a) Mode structure of the same GeSbS microresonator in $TE_{00}$ mode. Solid and dashed lines show the simulated integrated dispersion for $TE_{00}$ and high-order modes. Blue circles represent the measured integrated dispersion of the $TE_{00}$ mode. The overall mode displays normal dispersion, $D_2/2\pi$ = -8.23 MHz, while at $\mu = -4$ (corresponding to ca. 1543 nm), the dispersion changes locally to form the mode crossing. Inset: Zoom-in transmission spectrum at $\mu = -4$. (b) Normalized comb power when the laser was scanned from the blue side to the red side. (c) The measured output optical spectra of dark-pulse combs at different stages as indicated in (b), black line denotes the simulated spectrum in stage III. (d) Simulated time-domain intensity profile, showing a dark pulse property and the duty cycle of the solitary wave is ca. 27.4 %. (e) Low-intensity noise spectrum of the microcomb measured in a photodiode.

Moreover, given that the $TE_{00}$ mode family is in the normal dispersion regime, we further test the possibility of dark pulse-based comb generation in the GeSbS microresonator. Compared with the bright soliton comb in the anomalous dispersion regime, the dark-pulse comb could feature a higher conversion efficiency, while its generation usually requires a localized anomalous effect to initiate the sideband comb modes. By careful analysis of our microresonators, we notice that there exist several localized mode coupling positions in the mode spectrum of the resonator, as indicated by FEM mode calculations, by the coupling between the fundamental $TE_{00}$ mode (solid line) and high-order modes (dashed lines), see **Fig. 5a**. The mode coupling would lead to avoided mode crossings (AMXs) on the mode spectrum and introduce localized anomalous effect to the mode where the coupling occurs[1]. In this way, we may choose to pump an AMX mode for dark-pulse comb generation using the laser tuning scheme.

In experiment, the $TE_{00}$ mode of the microresonator shows an overall normal GVD of $D_2/2\pi = -8.23$ MHz, and the selected AMX mode is at 1543 nm. The comb generation was observed by manually tuning the laser from the blue side to the red side of the resonance when the pump laser was launched into the bus waveguide with 25 mW (see **Fig. 5b**). The comb power trace shows three steps during the laser tuning process, associated with the different states of the microcomb in normal dispersion region[53]. **Figure 5c** shows the evolution of the output optical spectra. At stage I, 3-FSR spaced comb lines were initially observed. Next to stage II, the bandwidth was increased with further tuning the pump frequency, and a flat wing at 1578 nm was formed. Finally, when the pump laser was stopped at stage III, we obtained the microcomb covering the range from 1510 nm to 1590 nm with a repetition rate of ca. 200 GHz. As a critical factor, the measure power conversion efficiency (CE = $P_{other}^{out}/P_{pump}^{in}$, where $P_{other}^{out}$ is the sum of all comb lines excluding the pump and $P_{pump}^{in}$ is the input power on-chip) is around 21.8%, which could be improved by optimizing the coupling structure or introducing coupled-microresonator geometry for dynamic adjustment of AMXs based on our GeSbS ChG platform[8].

Meanwhile, we performed numerical simulations based on the Lugiato-Lefever equation (LLE) model[54]. As a result, the simulated waveform is in good agreement with the experimental result, see **Fig. 5c and 5d**. In addition, the duty cycle of the temporal dark-pulse pattern is simulated to be around 27.4%, which is also close to the measured conversion efficiency[54], see **Fig. 5d**. Moreover, the comb was assessed on the low-frequency spectrum in the radio-frequency (RF) domain and showed a low-noise figure.

## Conclusion

We have presented an integrated nonlinear photonics platform for soliton microcombs based on our home-developed GeSbS microresonators, which featured high $Q$-factor up to ca. $2.3 \times 10^6$, higher nonlinearity, low TOC, and high LDT. The OPO pump threshold as low as 780 µW was attained. We achieved completely different dispersion curves by precise dispersion engineering, with $TE_{00}$ being normal dispersion and $TM_{00}$ being anomalous dispersion, respectively. Moreover, both a bright soliton-based microcomb and a dark-pulsed comb were realized in a single microresonator, in its separated fundamental polarized mode families with the ten-milliwatt pump power level. Our results pave the way that our homemade GeSbS is a robust and compact integrated nonlinear platform and reveals potential system-level applications by heterogeneously full-integrated on silicon, such as MIR frequency combs, and optical frequency synthesizers.

**Materials and Method**

**Chalcogenide film deposition.** High purity elements (99.9999% Ge, 99.9999% Sb and 99.9999% S) were used as the starting materials to prepare the $Ge_{20}Sb_{10}S_{65}$ glass by the conventional melt-quenching technique. The high purity $Ge_{20}Sb_{10}S_{65}$ glass was fabricated by modified physical and chemical purification techniques[55], which was used as the starting glass for ChG film. The thermal evaporation method was used to deposit the $Ge_{20}Sb_{10}S_{65}$ film on Si substrates with a 3 μm $SiO_2$ layer in a vacuum chamber at a base pressure of $7\times10^{-6}$ Pa. The substrates were mounted on a rotatable hold and pretreated using Ar plasma to improve the adhesion between the films and substrates. The evaporation rate was set to approximately 5-6 nm/min.

**Microresonator fabrication.** As illustrated in **Fig. 2a**, a layer of GeSbS (850 nm) was deposited on top of Si substrate with a 3 μm $SiO_2$ layer. The chip was put into a vacuum furnace and processed at 350 °C. It was then coated with a photoresist (ARP-6200) with a thickness of ca. 800 nm and patterned by electron beam lithography (EBL). After development, a thermal reflow process was applied on a hotplate at 140 °C for 3 mins to remove the roughness of the sidewalls of the patterns. RIE with $CF_4$/$CHF_3$/Ar gas chemistries was then applied to transfer the patterns to the GeSbS layer. Afterward, an ICP-RIE was used to remove the residual resist. Finally, the sample was clad with 3-μm-thick $SiO_2$ by ICP-CVD deposition at 300 °C. The inverse tapers and microresonator were simultaneously prepared under the same fabrication conditions, achieving the lowest coupling loss of -2.3 dB (45 %) per facet at 1550 nm.

**Characterizations of LDT, Q factor, dispersion-engineering, and comb generations.** The LDT was measured using a similar method[56] that the $Ge_{20}Sb_{10}S_{65}$ films were shined with a femtosecond laser from an optical parametric amplifier (Coherent, Opera-HP) with a repetition rate of 10 kHz, a center wavelength of 1550 nm, and a pulse width of 250 fs. The laser was focused on a spot size of 200 μm on the top surface of the ChG film by a $CaF_2$ lens with a focal length of 50 mm. The temperature was stabilized at 30 °C to minimize the temperature drift-induced resonance shift. A continuous-wave tunable laser (Keysight 81606A) was used to characterize the linear properties of microresonators.

To characterize the dispersion of GeSbS microresonators, a fiber ring resonator consists of an optical coupler with a splitting ratio of 90:10, and a fiber loop with a length of ~ 10 m was utilized to calibrate the frequency difference[21]. The probe light was divided into two channels (passing through the GeSbS resonator and the fiber ring resonator, respectively). Then, they were received by two photodetectors (New Focus, Model 1811FC) monitored by an oscilloscope (Keysight, DSOS404A) to record transmission traces.

For Kerr frequency combs generation, a C-band continuous-wave tunable laser (Toptica CTL 1550) was amplified by an erbium-doped fiber amplifier (EDFA) (Amonics, AEDFA-33-B-FA), the output light was split into three parts, see SI **Fig. S4**. One was recorded by an optical spectrum analyzer (OSA) (YOKOGAWA AQ6370D). The second was detected by a photodetector, which was monitored by an electrical spectrum analyzer (ESA) (Agilent N9030A) for determining the intensity noise of the generated comb. The third part was received by a photodetector, which was monitored by the oscilloscope for recording the power trace of pump light.

**Supporting Information**

Supporting information is available from the corresponding author upon reasonable request.

**Acknowledgments**

National Key R&D Program of China under Grant (2019YFA0706301), Key Project in Broadband Communication and New Network of the Ministry of Science and Technology (MOST) (2018YFB1801003), National Science Foundation of China (NSFC) (U2001601, 61975242, 61525502, 11974234), the Science Foundation of Guangzhou City (202002030103).

**Author contributions**

D Xia, P Zeng, Z Yang, J Wu, B Zhang and H Guo conceived the experiment, D Xia, Z Wang, L Luo, did the simulation and design for the devices. D Xia, J Wu, J Huang, Z Wang and B Zhang performed the experiments. P Zeng, Z Yang, Mi Gao, D Liu, and S Yang did the fabrication. D Xia P Zeng, Z Yang, J Wu, B Zhang and H Guo analyzed the data. B Zhang and Z Li supervised this project. All the authors contributed in writing the manuscript.

**Conflict of Interest**

The authors declare no conflict of interests.